# Detection of graphene's divergent orbital diamagnetism at the Dirac point

J. Vallejo Bustamante[1], N. J. Wu[1,6], C. Fermon[2], M. Pannetier-Lecoeur[2], T. Wakamura[1,7], K. Watanabe[3], T. Taniguchi[4], T. Pellegrin[1], A. Bernard[1], S. Daddinounou[1], V. Bouchiat[5], S. Guéron[1], M. Ferrier[1], G. Montambaux[1], H. Bouchiat[1]*

The electronic properties of graphene have been intensively investigated over the past decade. However, the singular orbital magnetism of undoped graphene, a fundamental signature of the characteristic Berry phase of graphene's electronic wave functions, has been challenging to measure in a single flake. Using a highly sensitive giant magnetoresistance (GMR) sensor, we have measured the gate voltage–dependent magnetization of a single graphene monolayer encapsulated between boron nitride crystals. The signal exhibits a diamagnetic peak at the Dirac point whose magnetic field and temperature dependences agree with long-standing theoretical predictions. Our measurements offer a means to monitor Berry phase singularities and explore correlated states generated by the combined effects of Coulomb interactions, strain, or moiré potentials.

Orbital magnetism results from the quantum motion of electrons in a magnetic field. At low energy, this motion leads to the Landau spectrum, which is, in most two-dimensional (2D) conductors, a harmonic oscillator–type spectrum with equally spaced levels separated by the cyclotron energy $\hbar\omega_c$ (1). As long as the material is nonsuperconducting, this spectrum causes a very small diamagnetic low-field susceptibility that is usually hidden by spin contributions. However, some materials, such as graphene, can display extraordinarily large diamagnetism. This was predicted in the theoretical work of McClure (2), who showed that graphene is diamagnetic at half filling (at the so-called Dirac point), with a divergent zero-field susceptibility (the derivative of the magnetization $M$ with respect to the magnetic field $B$),

$$\chi_0(\mu) = \frac{\partial M}{\partial B} = -\frac{2e^2 v_F^2}{3\pi} \delta(\mu) \qquad (1)$$

where $v_F$ is the Fermi velocity, $e$ is the electronic charge, and the Fermi energy $\mu$ is zero at the Dirac point. This is all the more surprising because the density of states is zero at that point. The reason for this singular susceptibility stems from the electron-hole symmetric linear spectrum of Dirac relativistic electrons, which gives rise to a Landau spectrum quantized as $\pm\sqrt{nB}$ where $n$ is a positive integer.

The diamagnetic sign of the response is attributable to the existence of the zero-energy Landau level ($n = 0$), as recalled and sketched below [see also figure 5 of (3) and related comment]. This peculiar level is known to result from the Berry phase (4) of $\pi$ acquired by the wave function pseudo-spin upon a revolution around a Dirac cone in reciprocal space (5). Therefore, the diamagnetic sign of the susceptibility at the Dirac point is a direct consequence of the $\pi$ Berry phase. Indeed, it has been shown that slightly different models with a zero Berry phase lead to orbital paramagnetism at the Dirac point (3). To summarize, the divergence reflects the linear spectrum and the diamagnetic sign reflects the nontrivial geometry of the eigenstates via the Berry phase (3).

However, despite these striking predictions, the singular orbital magnetism of a single graphene flake remains undetected. The reason for this lies in at least three obvious experimental challenges. First, the magnetic signal of an atomic monolayer is extremely small. Second, the McClure singularity, originally computed for an ideal system without disorder at zero temperature and in the limit of zero magnetic field, is rounded when any of these conditions is relaxed (6–9). Finally, this orbital magnetism is expected to be hidden by the magnetism of spins originating from edges, vacancies, or impurities (10), which tends to become dominant at low temperatures. This may explain why magnetization measurements have to date only been performed on a macroscopic number of graphene flakes. In one case (11), the focus was mainly on the spin paramagnetism of induced vacancy– and resonant states–type defects, which were found to depend on the chemical doping of the samples. A second set of measurements (12) did focus on the diamagnetism, and found a diamagnetism larger than that of pure graphite by a factor of 3. The magnetization curves at high fields were found to be compatible with the $\sqrt{B}$ dependence predicted for the Dirac spectrum. However, in those experiments it was not possible to fix the doping, nor could the residual contribution of paramagnetic spins along the edges of the flakes be well controlled (13).

In the present experiment, by contrast, we measure the orbital moment of a single flake whose Fermi energy is precisely controlled. This is achieved by implementing several sensitivity-enhancing features detailed in (14). As shown in Fig. 1, our experiment consists of a graphene monolayer, encapsulated between two hexagonal boron nitride (hBN) 2D crystals, capacitively coupled to a top-gate electrode and positioned above a highly sensitive magnetic detector made of two giant magnetoresistance (GMR) strips (figs. S1 to S3) in a Wheatstone bridge configuration. One key asset is that whereas graphene's orbital magnetism responds to a field perpendicular to the graphene plane ("vertical" field), the resistance of the GMR detectors only depends on the in-plane field, and thus detects the horizontal component of the field generated by the orbital current loops in the graphene (Fig. 1), all the while being insensitive to the applied vertical field. A second feature is the addition of a small AC modulation to the DC gate voltage, which in turn modulates the magnetization with respect to gate voltage and thus the resistance of the GMR detector. Beyond increasing the sensitivity, this modulation technique makes gate-independent magnetic signals invisible. Thanks to these experimental implementations, we were able to detect the derivative with respect to gate voltage of the diamagnetic McClure peak at low magnetic fields. We have also measured the crossover to the de Haas–van Alphen magnetic oscillations at higher fields.

Figure 2 shows the gate voltage derivative of the field induced by the graphene sample on the calibrated GMRs as a function of $V_g$ for perpendicular magnetic fields between 0.1 and 1.2 T. We found an antisymmetric peak centered at $V_g = -0.29$ V, which we identified as the Dirac point by comparing to the position of the maximum in the resistance of the sample $R(V_g)$ (Fig. 2B and fig. S3). At low magnetic fields, the antisymmetric peak detected in the GMR resistance is directly proportional to the derivative of the McClure peak with respect to the chemical potential (controlled by the gate voltage), as detailed in (14). The experimental detection of this peak and its evolution with magnetic field are the central result of our work. Both the peak width and amplitude increase linearly with field, as shown in Fig. 2, E and F. Above 0.6 T, $\partial M/\partial V_g(V_g)$ displays periodic oscillations in addition to the antisymmetric

[1]Université Paris-Saclay, CNRS, Laboratoire de Physique des Solides, 91405 Orsay, France. [2]SPEC, CEA, CNRS, Université Paris-Saclay, 91191 Gif-sur-Yvette, France. [3]Research Center for Functional Materials, National Institute for Materials Science, 1-1 Namiki, Tsukuba 305-0044, Japan. [4]International Center for Materials Nanoarchitectonics, National Institute for Materials Science, 1-1 Namiki, Tsukuba 305-0044, Japan. [5]Néel Institute, CNRS, 38000 Grenoble, France. [6]Université Paris-Saclay, CNRS, Institut des Sciences Moléculaires d'Orsay, Orsay, France. [7]NTT Basic Research Laboratories, NTT Corporation, Atsugi, Kanagawa, Japan.
*Corresponding author. Email: helene.bouchiat@u-psud.fr

**Fig. 1. Experimental setup.** (**A**) Principle of the experiment. The orbital magnetization $M_{orb}$ can be viewed as a current loop (blue circle) generated by a vertical magnetic field $B$ and circulating around the graphene region covered by the gate electrode. It is detected by the two GMR detectors, which measure the horizontal components $B_1$ and $B_2$ (respectively on the detectors $GMR_1$ and $GMR_2$) of the magnetic field (black dashed lines) generated by this loop. The sensitivity is on the order of 0.1 nT (*14*). (**B**) Micrograph of the sample investigated; the gate voltage derivative of the orbital magnetization is measured via the difference between the DC current–biased $GMR_1$ and $GMR_2$ resistances with $I_1$ and $I_2$ adjusted so as to cancel the DC component of the voltage difference $V_1 - V_2$. The signal measured by a lock-in amplifier (L.I.) is the AC component of $V_1 - V_2$ at the modulation frequency of the gate voltage. There is no current applied to the graphene sample during the magnetization measurements.

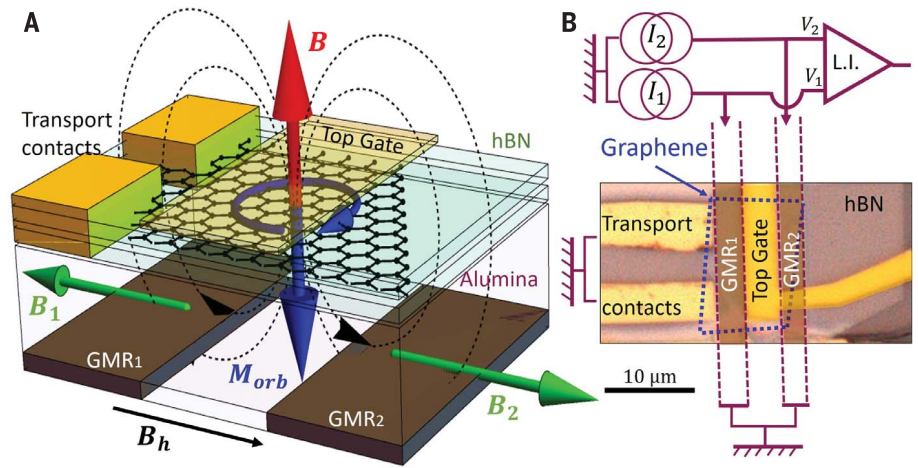

**Fig. 2. Magnetization data.** (**A**) Detected modulation of the GMR detector's resistance with an AC gate voltage modulation of 20 mV, as a function of the DC gate voltage $V_g$. The quantity plotted is $\partial B_{GMR}/\partial V_g$, where $B_{GMR}$ is deduced from the signal on the calibrated GMR sensor divided by the applied vertical magnetic field $B$. Data are the average of 80 independent measurements. (**B**) Derivative with respect to gate voltage of the two-point resistance of graphene measured through the side electrodes, in the region of the Dirac point, with a gate voltage modulation of 50 mV. (**C**) For comparison, the GMR signal at −0.6 T using the same gate voltage modulation as in (B). The GMR peak is much narrower.

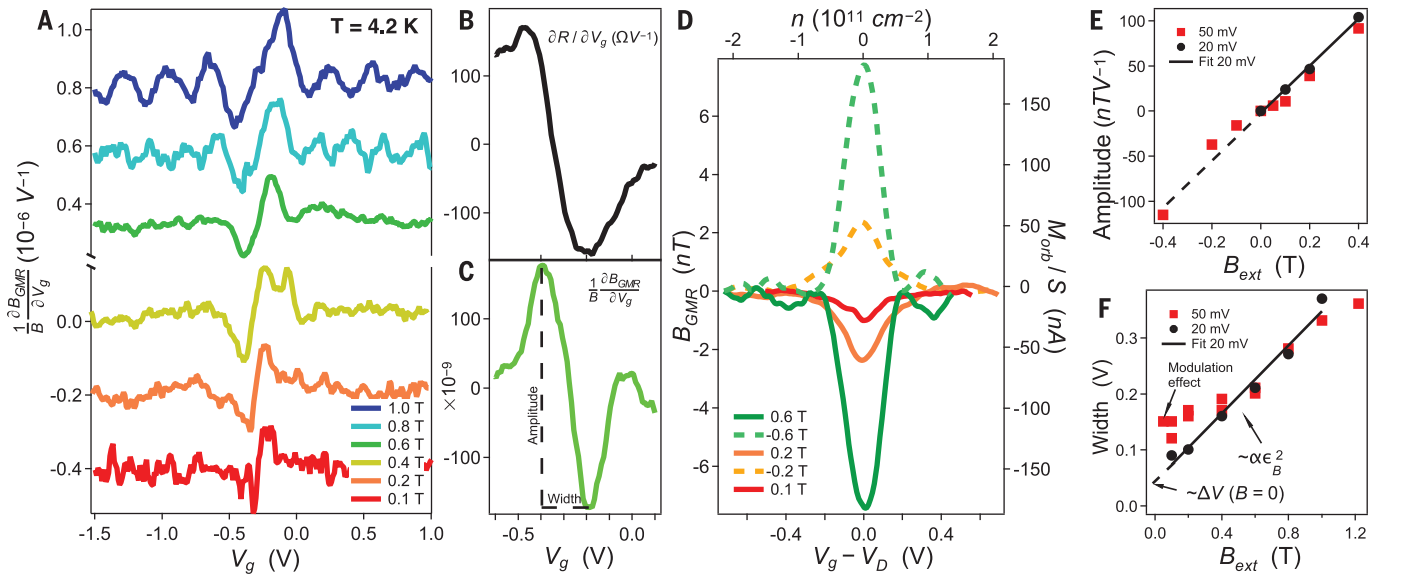

(**D**) Numerical integration of the data plotted in (A) and fig. S4, yielding the magnetization per unit surface (in nA; right axis) and the magnetic field $B_{GMR}$ detected by the GMR device (in nT; left axis) as a function of the gate voltage. (**E** and **F**) Field dependences of the GMR peak maxima and widths, as defined in (C), for gate voltage modulations of 20 mV (circles) and 50 mV (squares), and comparison with the linear variations expected theoretically (see Eqs. 5 and 6 and eqs. S20 to S27), using the scaling between the gate voltage and the square of the Landau energy $\varepsilon_B^2$ via the parameter $\alpha$ defined in Eq. 8. Deviations from linearity caused by excessive modulation amplitudes are visible for a 50-mV modulation.

peak around the Dirac point. These oscillations are related to the expected de Haas–van Alphen oscillations of the magnetization, as discussed below.

The magnetization, shown in Fig. 2D, is obtained by the integration of the curves in Fig. 2A. The peak amplitude translates into a few nanoteslas induced in the GMR plane by graphene's orbital response to a 0.1-T per-pendicular field. This illustrates the sensitivity of our experiment. The correspondence between this detected field, $B_{GMR}$, and magnetization is obtained by modeling the orbital magnetic moment as an effective current loop whose geometry is defined by the gated region of graphene (fig. S10). We find that positive magnetic fields produce a negative peak in magnetization, and vice versa, which is con-sistent with the expected diamagnetic response of graphene (*2*). The sign of the response was carefully determined via the sign of the response of the GMR sensor to a horizontal field of known orientation. We can assert that the signal cannot be attributed to gate voltage–dependent magnetism of paramagnetic impurities, given the absence of temperature dependence between 4.2 and 40 K (*15*) (see fig.

S9). In addition, thanks to our gate modulation technique, we can exclude spurious contributions from impurities or defects in alumina or graphene, which would not depend on gate voltage. This contrasts with all previous measurements of graphene's magnetism, which were performed on large ensembles of flakes.

In the following, we compare our results to theoretical predictions, taking into account the variations of the chemical potential caused by charge inhomogeneity, and ignoring the smaller broadening due to temperature (14). Assuming a Gaussian distribution for the electrochemical potential $\mu'$ of standard deviation $\sigma$,

$$P_\sigma(\mu') = \frac{1}{\sqrt{2\pi}\sigma} \exp\left(-\frac{\mu'^2}{2\sigma^2}\right) \quad (2)$$

yields a smoothed susceptibility,

$$\chi_\sigma(\mu) = \int P_\sigma(\mu')\chi_0(\mu - \mu')d\mu' \quad (3)$$

Then, the $\delta$-peak of the susceptibility is broadened as

$$\chi_\sigma(\mu) = -\frac{2e^2 v_F^2}{3\pi} P_\sigma(\mu) \quad (4)$$

The full field and chemical potential dependence of the magnetization, including the oscillations, is given by the derivative $M = -\partial\Omega/\partial B$ of the disorder-averaged grand potential $\Omega_\sigma(\mu, B)$ (14) (eqs. S20 to S23):

$$\Omega_\sigma(\mu, B) = \int P_\sigma(\mu')\Omega_0(\mu - \mu', B)d\mu' \quad (5)$$

with

$$\Omega_0(\mu, B) = \frac{\varepsilon_B^3}{4\pi^2\hbar^2 c^2} \sum_{p>0} \frac{1}{p^{3/2}}\left[1 - 2S\left(2\sqrt{p}\frac{|\mu|}{\varepsilon_B}\right)\right] \quad (6)$$

The Landau levels at energies $\sqrt{n}\varepsilon_B$, with $\varepsilon_B = \sqrt{2e\hbar v_F^2 B}$, enter via the argument $\sqrt{p}|\mu|/\varepsilon_B$ where $p$ is an integer, in the Fresnel function

$$S(x) = \int_0^x \sin\frac{\pi}{2}t^2 dt \quad (7)$$

The predicted disorder-averaged magnetization is displayed in Fig. 3E. With increasing field, it evolves from a sole diamagnetic McClure peak of width $\sigma$ to a broader peak with additional oscillations, centered at $\mu_n/\varepsilon_B = \sqrt{n}$. Figure 3D demonstrates how charge disorder induces rounding and attenuates the oscillations.

To compare these predictions to experiment, we must also relate the gate voltage $V_g$ to the chemical potential $\mu$. Far from the Dirac point, this relation is quadratic, $V_g(\mu) - V_D = \alpha\mu^2$ sign($\mu$), with

$$\alpha = \frac{e/C_g}{\pi\hbar^2 v_F^2} \quad (8)$$

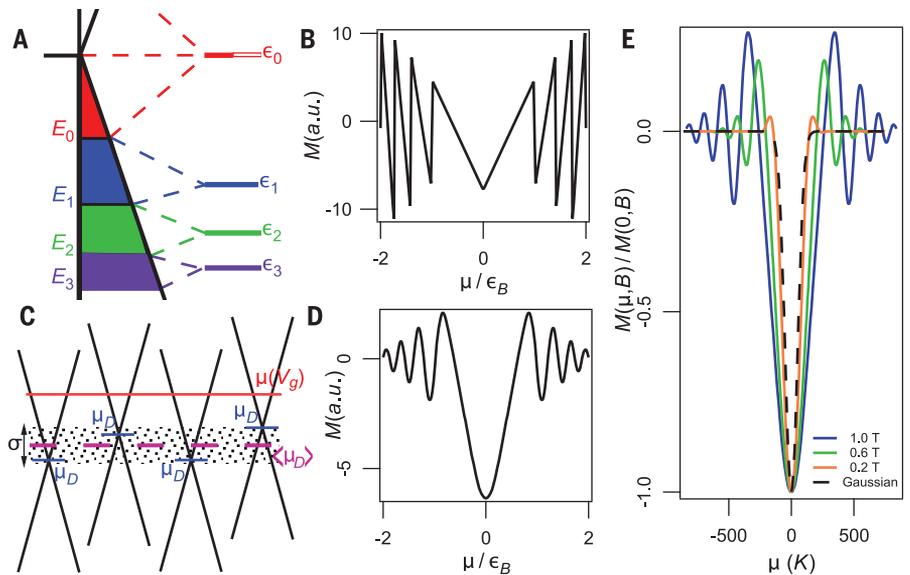

**Fig. 3. Calculated chemical potential dependence of the orbital magnetization of graphene in a finite magnetic field.** The calculations are based on Eq. 6; see (14) for more details. (**A**) Evolution of the graphene spectrum in a magnetic field [adapted from (3)]. The condensation of the continuous spectrum into Landau levels decreases the energy, except for the zero energy level whose contribution is predominant. Globally, the net result is an increase of the energy with the field—that is, a diamagnetic response [see also figure 5 of (3)]. (**B**) Without disorder, the magnetization, plotted as a function of the rescaled chemical potential $\mu/\varepsilon_B$, exhibits discontinuities at the Landau level energies $\sqrt{n}\varepsilon_B$; a.u., arbitrary units. (**C**) Sketch illustrating the spatial distribution of electrochemical potentials $\mu' = \mu_D - \langle\mu_D\rangle$ where $\mu_D$ is the local Dirac point and $\langle\mu_D\rangle$ is its spatial average. (**D**) Rounding of $M(\mu/\varepsilon_B)$ by a Gaussian chemical potential distribution with a variance $\sigma = 0.1\varepsilon_B$. (**E**) Calculated $M(\mu)$ for different magnetic fields for $\sigma = 50$ K. At low fields, the oscillations disappear and the magnetization displays a Gaussian diamagnetic peak at $\mu = 0$. This peak is broadened by the magnetic field as soon as $\varepsilon_B \geq \sigma$.

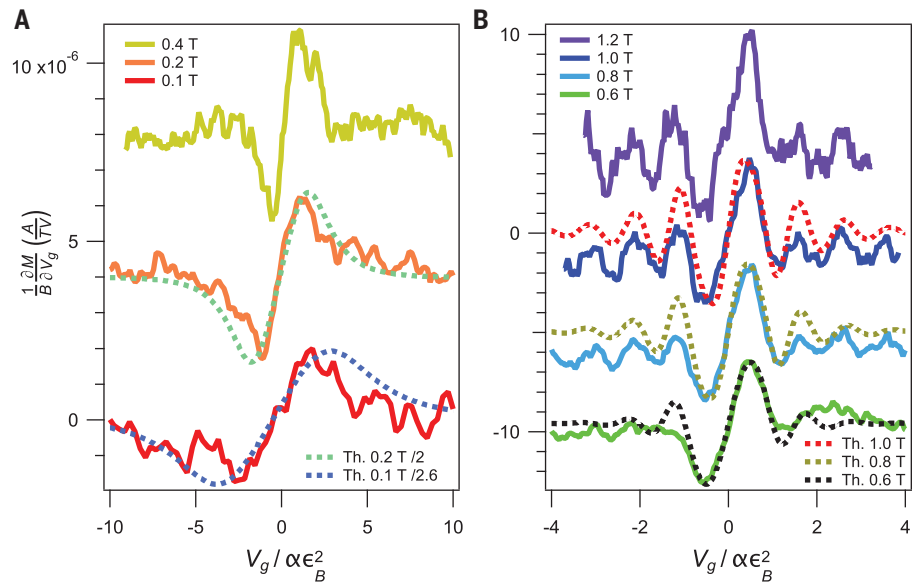

**Fig. 4. Comparison of theory to experiment.** Fit of detected AC magnetization response to a gate voltage modulation of 50 mV, as a function of the DC gate voltage divided by $\alpha\varepsilon_B^2 = 2\alpha e\hbar v_F^2 B$. Dashed lines show the theoretical gate dependence of $\partial M/\partial V_g$, with $\sigma_0 = 165$ K and $\sigma_\infty = 50$ K, including the extra rounding effect owing to the 50-mV AC gate modulation. In (**A**), the amplitude of the theoretical signal has been rescaled by a factor of 1/2.6 at 0.1 T and by a factor of 1/2 at 0.2 T to fit quantitatively the experimental data. In (**B**), the rescaling factors are closer to unity for higher fields for which the McClure peak is expected to be independent of $\sigma_0$.

where $V_D$ is the gate voltage at the Dirac point, and $C_g$ is the geometrical capacitance per unit surface between graphene and gate, as determined from the $V_g$ periodicity of the de Haas–van Alphen oscillations at high field (14). In contrast, close to the Dirac point, $V_g$ varies linearly with $\mu$, with a slope given by the standard deviation of $\mu$ disorder around the Dirac point, $\sigma_0$:

$$V_g(\mu) - V_D = \frac{4\sigma_0\mu}{\sqrt{2\pi}} \qquad (9)$$

(eqs. S24 and S25). We find that the experimental data can be fit (see Fig. 4) using two constants, $\sigma_0 = 165$ K and $\sigma_\infty = 50$ K, which describe the $\mu$ distribution at low and high doping respectively (eqs. S26 and S27). The smaller value of $\sigma_\infty$ is explained by the more efficient screening of charge impurities at high doping. We note that the two constants can practically be determined independently, given the high sensitivity of the decay of the de Haas–van Alphen oscillations to disorder, and the large broadening of the McClure peak induced by magnetic field (on the order of $\varepsilon_B$).

We find that $M(V_g)$ and $\partial M/\partial V_g$ depend on $V_g$, $\sigma_0$, and $\sigma_\infty$, exclusively via the variables $V_g/\alpha\varepsilon_B^2$, $\sigma_0/\varepsilon_B$, and $\sigma_\infty/\varepsilon_B$. In particular, the variation as $\alpha\varepsilon_B^2$ of the $\partial M/\partial V_g$ peak's width, shown in Fig. 2F, is directly related to this scaling, which originates from the Dirac Landau spectrum of graphene.

Next, we compare the magnetization peaks measured at the Dirac point at 0.1 T and 0.2 T to theoretical expectations. We find that the predicted amplitude of the antisymmetric magnetization peak at the Dirac point ($1/B$) ($\partial M/\partial V_g$) at low magnetic field, equal to $9.6 \times 10^{-6}$ $A(TV)^{-1}$, is on the order of the experimental values, although larger by a factor 2 to 2.6. This value corresponds to a diamagnetic magnetization two orders of magnitude larger than the Landau diamagnetism of a 2D free electron gas. Finally, deviations from the linearity between magnetization and magnetic field are expected when $\varepsilon_B$ becomes much greater than $\sigma_0$, with a smooth crossover toward a $\sqrt{B}$ dependence (eqs. S20 to S23). Because the calibration of the GMR sensor becomes delicate in high perpendicular magnetic fields owing to the residual imperfect alignment of the magnetic field, these deviations from linearity cannot be precisely checked in the field range above 0.5 T where they are expected to occur.

We have detected the McClure singularity of low-field orbital magnetization of a single graphene monolayer at the Dirac point, which is the signature of the $\pi$ Berry phase of electronic wave functions in graphene. This experiment should also enable the investigation of interband-induced Berry curvature anomalies (16–18) as well as Coulomb interaction effects in 2D materials such as graphene and its bilayer (19, 20). Moreover, in contrast to the diamagnetic McClure peak observed here, a divergent paramagnetic orbital susceptibility (21) is expected at Van Hove singularities in the presence of moiré potentials of high periodicity. These moiré potentials also generate flat bands in the magic-angle twisted bilayer of graphene (22). An anomalous quantum Hall effect is then expected to appear as the result of Coulomb interactions leading to valley symmetry breaking (23–25) and orbital current loops in zero magnetic field. They are detectable via the orbital magnetic moments they would generate, as very recently shown in (26). The possibility of generating flat bands with a periodic array of strain has also been predicted (27–29). In fig. S12, we present data on a strained sample on which it was possible to detect a gate-dependent GMR signal at zero magnetic field. This preliminary result suggests that more controlled situations like that in (30) can be investigated. Such measurements could also be used to reveal the expected ballistic loop currents along the edges of 2D topological insulators (31–34).

## ACKNOWLEDGMENTS

We thank E. Paul of SPEC-CEA for the GMR sensors patterning, and R. Deblock, A. Chepelianskii, F. Piéchon, J. N. Fuchs, F. Parmentier, R. Delagrange, A. Murani, and S. Sengupta for fruitful discussions. **Funding:** Supported by the BALLISTOP ERC 66566 advanced grant and the MAGMA ANR-16-CE29-0027-01 grant. Also supported by the Elemental Strategy Initiative conducted by the MEXT, Japan, grant JPMXP0112101001. JSPS KAKENHI grant JP20H00354, and CREST (JPMJCR15F3), JST (K.W. and T.T.). **Author contributions:** J.V.B. fabricated and positioned the graphene stack on the GMR device, optimized and ran the experiment, and worked on the interpretation and fits of the data. N.J.W. and T.W. helped on sample fabrication. K.W. and T.T. provided hBN single crystals. T.P., A.B., S.D., and V.B. contributed to the design, optimization, and calibration of the experiment. C.F. and M.P.-L. designed, fabricated, and optimized the GMR sensors. G.M. is responsible for the theoretical part of the work. M.F., S.G., and H.B. supervised the experimental work. J.V.B., T.W., C.F., M.P.-L., G.M., S.G., and H.B. contributed to the writing of the manuscript. **Competing interests:** There are no competing interests. **Data and materials availability:** Datasets and theory curves computed using mathematica are available on Zenodo (35).

## SUPPLEMENTARY MATERIALS

science.org/doi/10.1126/science.abf9396
Materials and Methods
Supplementary Text
Figs. S1 to S17
References (36–44)

10 December 2020; resubmitted 14 August 2021
Accepted 19 October 2021
10.1126/science.abf9396

# Supplementary Materials for

## Detection of graphene's divergent orbital diamagnetism at the Dirac point

J. Vallejo Bustamante *et al*.

**The PDF file includes:**

# I. MATERIALS AND METHODS

## A. Measuring orbital magnetization of graphene with a GMR detector

### 1. The GMR detector

The magnetoresistive material of the GMR detectors consist in a multilayer stack of ferromagnetic layers.These layers are sufficiently thin to ensure negligible leakage fields. The low field magneto-resistance is determined by the in-plane orientation of the soft layer's magnetization. This variation is linear with the horizontal magnetic field $B_H$ on a few gauss scale around zero and saturates above a few mT.

*a. GMR effect* The Giant MagnetoResistance (GMR) effect [36] is based on the variation of conductivity in multilayered ferromagnetic materials according to the relative orientation of the different layer's magnetization. It is widely used for hard disk drive read heads or magnetic memories, as well as for sensing purposes. In this latter case, the spin valve structure is the most commonly used [37]. It comprises a magnetic layer exhibiting a strong coercivity (hard layer) separated from a magnetic layer with a very low coercivity (soft layer) by a thin metallic spacer. The magnetization of the soft layer can align along an in-plane applied field, whereas the direction of magnetization is fixed in the hard layer . The resistance of the whole stack varies with the angle between the magnetization directions of the two layers.

*b. GMR fabrication process* The GMR stack is deposited by sputtering (Rotaris deposition chamber from Singulus) on a 300 $\mu$m-thick silicon substrate insulated by a SiO$_2$ layer of 1 $\mu$m. It has the following composition: Ta (3 nm) / NiFe (5 nm) / CoFe (2.1 nm) / Cu (2.9 nm) / CoFe (2.1 nm) / Ru (0.85 nm) / CoFe (2nm) / IrMn (7.5 nm) /Ru (0.4 nm) / Ta (5 nm). The hard layer is composed of the antiferromagnet IrMn coupled to a synthetic ferromagnet (CoF/Ru/CoFe) whereas the soft layer is made of the CoFe/NiFe bilayer. After the GMR stack deposition, an oven annealing step at 1 T and 180 °C is performed to set the magnetization direction of the hard layer. The GMR sensors ( 2 $\mu$m wide and 20 $\mu$m long) are defined by optical lithography and ion beam etching. Contacts consist of a Ti(10 nm)/Au (100 nm) bilayer deposited by evaporation. The resistance of the contacted GMR is 150 Ohms. The sample is then protected by a 900 nm-thick Al$_2$O$_3$ passivation layer, deposited by sputtering.

The GMR sensor is sensitive to the component of the magnetic field in the plane of the sensor. When a strong magnetic field is applied out-of-plane with perfect alignment, there is a loss of sensitivity as the field increases. Indeed, the magnetization of the soft layer tends to rotate of the plane, resulting in a reduction of the component projected in the plane [38].

*c. Sensitivity of the GMR sensor* The field response of the sensor is characterized by the derivative $dR/dB_H$= 2.5 $\Omega$/mT around $B_H = 0$ which is optimum when $B_H$ is in-plane and aligned perpendicularly to the long dimension of the GMR ribbons, see Fig. S1-A . The magnetization measurements were performed between T=4.2 K and 70 K without any bias current through the graphene. The vertical magnetic field (perpendicular to the graphene plane) is created by a superconducting solenoid. In order to compensate for the inevitable misalignment between the vertical direction and the normal to the sample plane, two Helmholtz pairs of superconducting coils were used to precisely cancel the component of the field, in the GMRs plane.

Although the field sensitivity decreases a mentioned above, the GMR is still operational at vertical fields up to 0.8 T at room temperature and up to 1.2 T at low temperatures, see Fig. S1-B. It was important to recalibrate the GMR with the horizontal field (created by the Helmholtz coils) for each value of the vertical field. We checked this calibration both before and after each experiment. When changing magnetic field we tried to maintain as far as we could the horizontal component of the applied magnetic field close to zero in order to avoid saturation of the GMR during the process and possible hysteresis which would change their sensitivity.

*d. Limit of magnetic field detection* The magnetoresistive sensors are inserted into a Wheatstone bridge circuit with adjustable dc currents (in the 0.1 to 1mA range) through the 2 GMR strips, in such a way that the bridge voltage is zero in a uniform horizontal magnetic field. The bridge is read by a low noise voltage amplifier. We use a low noise voltage amplifier which input noise voltage is 2nV/sqrt(Hz) above 40Hz (the current noise of the order of $10^{-14} A/\sqrt(Hz)$ yield a negligible contribution through the 150 Ohms GMR sensors). We show in Fig. S2 the voltage noise of the amplifier together with the noise measured on the dc current biased GMR sensors. This data show that the amplifier noise is negligible compared to the intrinsic noise of the GMR We have also checked that the graphene signal on the GMR does not depend on frequency between 7Hz and 125Hz but is more noisy at low frequency because of the low frequency 1/f noise of the GMRs. From this figure it appears that it could have been also interesting to work at even larger frequency, see Fig. S3.However we are concerned with the fact that at high frequency the modulation of the gate voltage induces current in graphene giving rise to an out-of-phase signal on the GMR sensors. By limiting the frequency below 200 Hz, we ensure that this contribution is negligible compared to the in-phase component due to the equilibrium orbital moment of graphene. The lowest detectable magnetic field is then simply related to the voltage noise measured on the GMR bridge $V_n = 4nV/\sqrt{(Hz)}$ knowing the sensitivity of the GMRs $s(B_{perp})$ in Ohms/T and the dc current drive $I_{GMR} = 0,5mA$ which is the maximum allowed value to keep the graphene device at 4.2K. When data is averaged over time $\tau$ the sensitivity reaches $\delta B = V_n/(2sI_{GMR}\sqrt{2\pi\tau}) = 80$ pT for $\tau = 300s$ at $B_{perp} = 0.1T$ and 0.4 nT at $B_{perp} = 1T$ taking into account of the decrease of the sensitivity of the GMRs with perpendicular magnetic field.

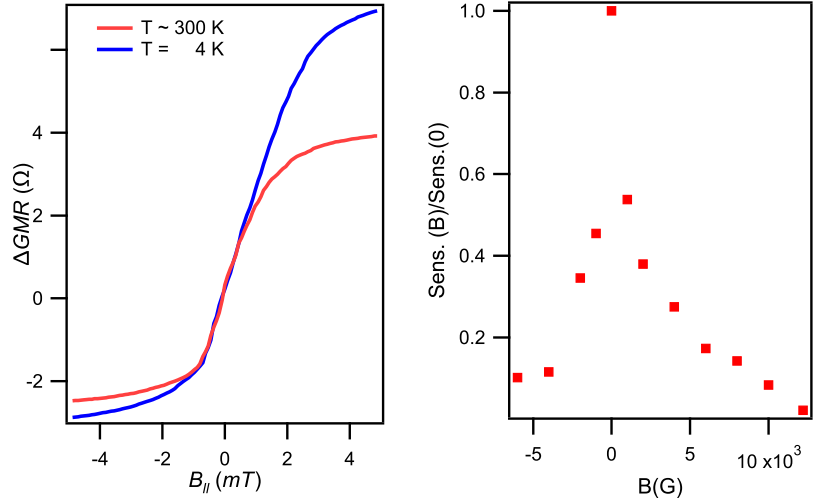

Figure S1: A-Magnetoresistance of one of the GMR resistances used in the experiment as a function of the in-plane magnetic field applied perpendicularly to the GMR ribbon. B-Relative sensitivity of the GMR detector as a function of vertical field. The maximum value ( without vertical field) corresponds a sensitivity of 2.5 $\Omega/mT$

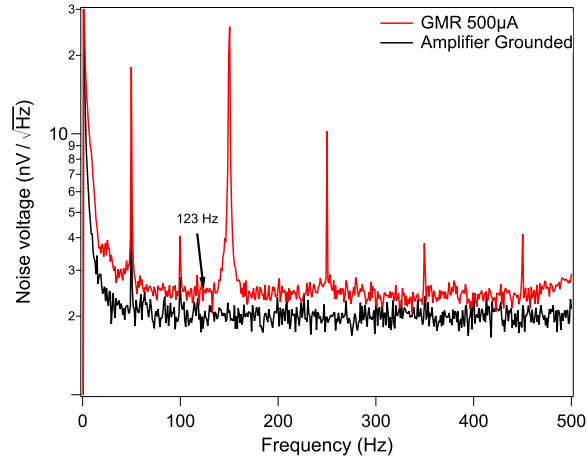

Figure S2: Noise spectrum of the shrt circuited GMR probes and with $500\mu A$ through the GMRs (red). The peaks at odd multiple values of 50Hz come from the power supply of the building sector.

TMR detectors are in principle more sensitive than the GMR sensors we have used. However they are much more noisy than GMRs in the range of temperature and frequency we are working at. Moreover the technique of fabrication of these sensors with the constraint that they also should work at cryogenics temperature with a small electric power consumption in order not to heat the graphene sample, is not as advanced as for the metallic GMR sensors we used.

It is also interesting to compare this sensitivity of detection of our GMR set up to SQUID detectors. If one considers a SQUID of 10 micron square 0.1 nT leads to a flux of the order of $10^{-21}W = 0.510 - 6\Phi_0$ which is of the order of the best sensitivity which can be achieved in zero magnetic field in a micro SQUID. Nano SQUIDs are even more sensitive in terms of flux detection. Their sensitivity in terms of magnetic field is however limited to 1nT, see D. Vasyukov, et al., Nat. Nanotech. 8 (2013) 639–644. They have a much better spatial resolution than our GMR detectors which translates into a higher sensitivity in terms of magnetic moment Y. Anahory et al. Nanoscale 12, 3174 (2020). They can now detect a single Bohr magneton electronic magnetic moments, however all these SQUID detectors are also very sensitive to out-of-plane magnetic field. In the present experiment we exploit the fact that GMR sensors are insensitive to the out-of-plane component of the magnetic field which is obviously not the case for SQUID sensors. NV sensors are very nice quantum detectors of small magnetic moments but they cannot be operated in magnetic fields above 100mT moreover they are not sensitive to the sign of the magnetic field detected, whereas this sign issue is essential in our experiment.

## 2. Assembling Graphene and detector

The graphene sample consists of a single monolayer flake encapsulated between hexagonal boron nitride (h-BN) flakes. We use dry transfer to assemble [39] the graphene stack and deposit it on the GMR detectors covered by a layer of alumina, see Fig. 1. A topgate electrode was deposited above the graphene region between the 2 magnetoresistive sensors. The graphene sample was subsequently contacted by 2 electrodes on one side of the magnetoresistive sensor after locally etching the top BN flake. The mobility of graphene $\mu_e$ in the vicinity of the Dirac point was estimated from the gate voltage dependent conductance $\partial G/\partial V_g = S\mu_e C_g$ and the capacitance $C_g$, to be of the order of 40 000 cm$^2$V$^{-1}$s$^{-1}$. This kind of experiment where it is possible to measure resistance and magnetization on the same sample in a wide range of magnetic field while controlling the chemical potential is quite challenging. We note that similar combined investigations of transport and magnetization were done on GaAs 2D electron gas systems [40], but only at large magnetic field . This possibility was essential in the present experiment for the determination of the Dirac point but also to check, by measuring the low field magnetoresistance, that the GMR detector does not perturb the magnetic field seen by the sample and finally to control possible heating effects induced by the current applied through the GMR sensors.

The magnetoresistive detector was calibrated in a test experiment with a current loop whose dimensions are similar to the gated region of graphene (8 $\mu$m x 4 $\mu$m). The orbital magnetization of graphene per unit surface expressed in current units can be obtained from this calibration and agrees with the calculation within the current loop model detailed below.

### B. Extra experimental information and data

#### 1. Resistance of the graphene sample

We show below the 2 wires measurement of the resistance of the graphene sample. The small amplitude of the resistance peak at the Dirac point is due to the small area of the gated graphene region.

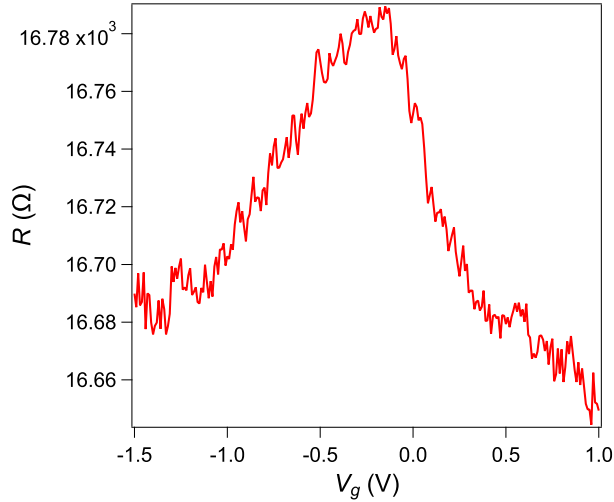

Figure S3: Resistance as a function of the gate voltage. The Dirac peak is located at the same gate voltage as the McClure peak shown in the main text.

#### 2. GMR data in positive and negative magnetic field

#### 3. Calculation of the thickness of the top BN flake

The oscillations observed in Fig.2 of the main paper at large gate voltage are manifestations of the de Haas-van Alphen effect. They are due to the contribution of non zero Landau levels to the magnetization.

The gate voltage capacitance can be simply deduced from the periodicity of these oscillations away from the Dirac point, corresponding to the region where $V_g = \alpha \mu^2$, with $\alpha = e/(\pi C_g \hbar^2 v_F^2)$. In particular the difference in the position $(V_g)$, between peaks 2 and 3 yields the geometrical capacitance between the gate electrode and graphene:

$$\Delta V_g = V_{g3} - V_{g2} = \alpha(\mu_3^2 - \mu_2^2) \tag{S.1}$$

$$\begin{aligned}
\Delta V_g &= \alpha(\varepsilon_B^2[N] - \varepsilon_B^2[N-1]) \\
&= \alpha(2e\hbar v_F^2 B)(N - N + 1) \\
\Delta V_g &= \frac{2e^2 B}{\pi C_g \hbar}
\end{aligned} \tag{S.2}$$

Recalling that $C_g = \varepsilon_0 \varepsilon_r/d$, being $d$ the BN thickness, we arrive to the expression:

$$d = \frac{\Delta V_g \pi \hbar \varepsilon_0 \varepsilon_r}{2e^2 B} \simeq 7 \times 10^{-8} \mathrm{m} \tag{S.3}$$

where we have taken $\Delta V_g = 0.32$ V (from the average of the peaks, $B = 1$ T and $\varepsilon_r = 3.8$ for BN. This rather large thickness of BN ensures that we can neglect the effect of the quantum capacitance in series with $C_g$.

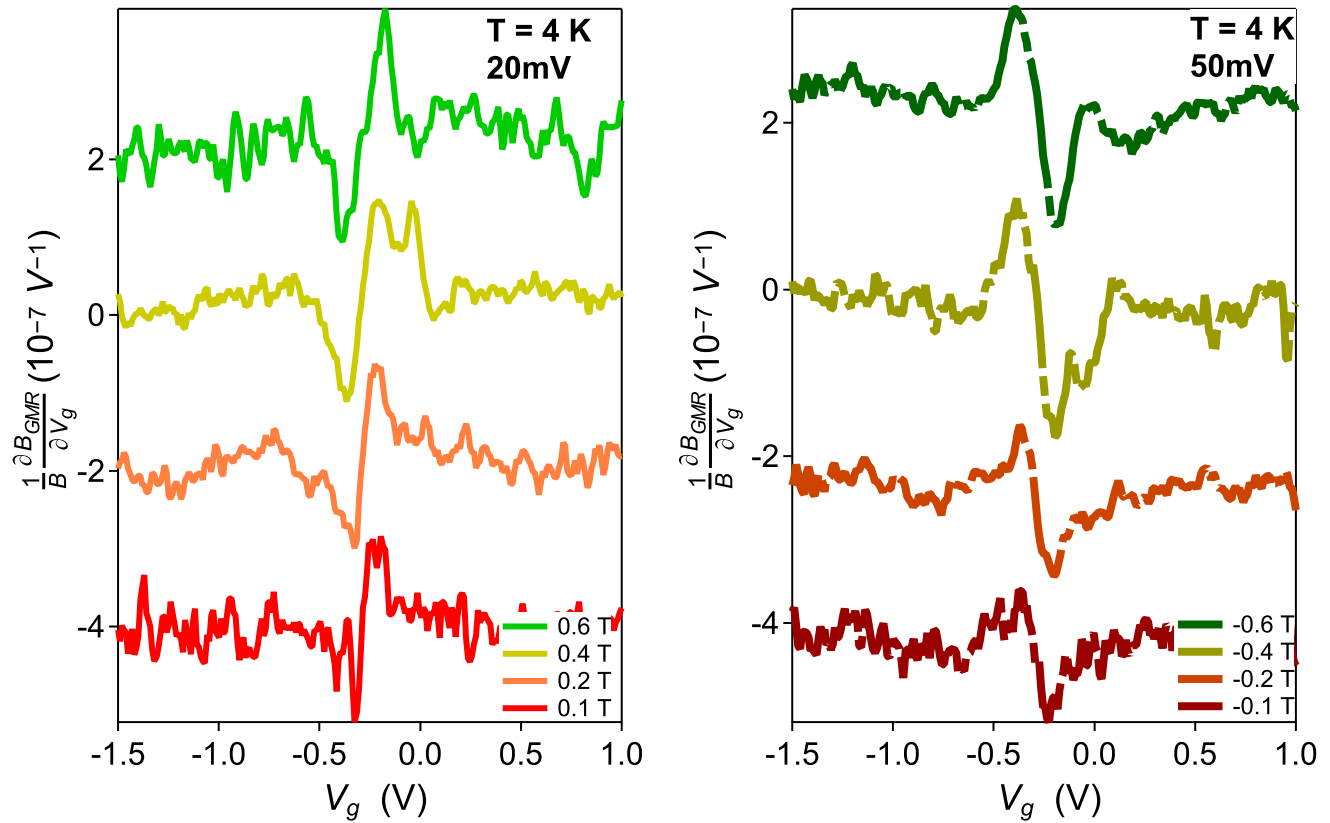

Figure S4: GMR Signal at positive and magnetic fields. This figure corresponds to the data obtained on the gate voltage modulated GMR resistance measured both for positive and negative values of magnetic field and leading to the data presented in Fig.2D after integration with respect to the gate voltage.

### 4. Influence of the amplitude modulation

The modulation $\delta V_g$ applied to the gate voltage, depending on its amplitude, can lead to a rounding of the magnetization peak and a reduction of the peak amplitude. It is necessary to find the good compromise between a small enough modulation amplitude to avoid this rounding and a large enough amplitude to preserve a good signal-to-noise ratio. In Fig. S4we plot the signal obtained on the GMRs, $\delta M$ renormalized by the modulation amplitude $\delta V_g$ for several values of $\delta V_g$. The amplitude of the signal measured is increased by a factor $\sim 3$ when $\delta V_g$ is decreased a factor 10. This effect is also visible in the integrated peak, M, in Fig. S5On the other hand we see in Fig. S8 that decreasing the modulation below 20mV does not lead to a detectable narrowing of the peak but only to more noisy data.

### 5. Temperature dependence

From the Gaussian model described in section 1 of these SM, we find that the value $\sigma = 50$ K reproduces quite well the damping of the de Haas-van Alphen oscillations seen at high magnetic field. This gives an approximate value below which, the effect of temperature should not be noticeable. One unavoidable source of heating in this experiment is the current through the GMRs. In Fig. S8 we can see that a change in the current in the GMRs by a factor 5 does not change the signal obtained. This is evidence that heating produced by the GMR is negligible in this range of current bias through the GMR sensor.

In Fig. S9 however, a decrease of the amplitude of the peak at $\mu = 0$ is clear when the sample was heated at 60 K. In addition, the de Haas-van Alphen oscillations start to disappear.

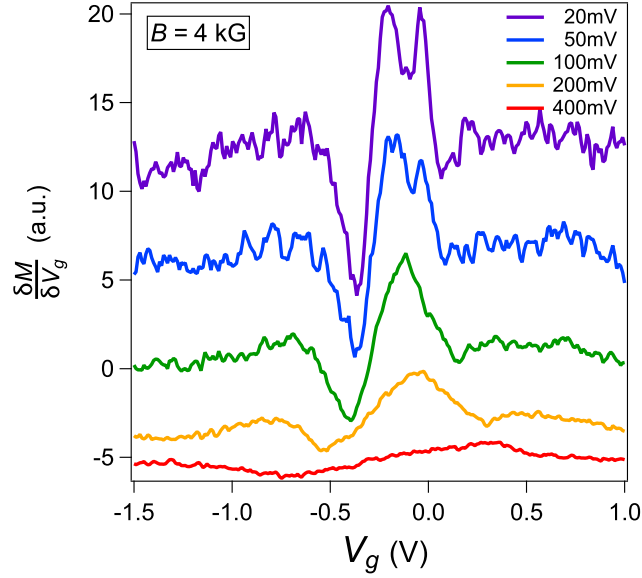

Figure S5: GMR signal $\delta M$ renormalised by the modulation amplitude $\delta V_g$ measured at $B_{perp} = 4$ kG, for different values of $\delta V_g$. The rounding effect of the amplitude is higher for high modulation amplitudes. Note also the splitting of the McClure peak at small modulation. This splitting is the signature of puddles which electrochemical potential lie outside of the main Gaussian peak.

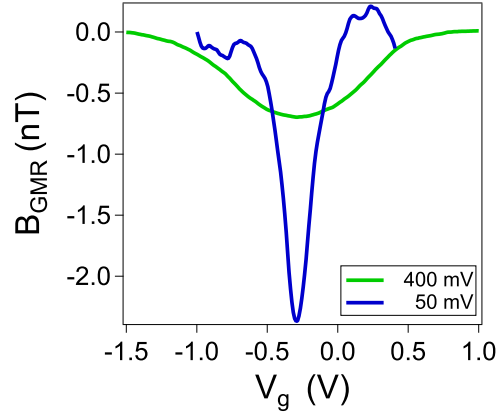

Figure S6: Integrated magnetization for an external field $B_{perp} = 2$ kG. In green, the data for 400 mV and in blue, data for 50 mV modulation.

### 6. Current loop model: from $V_{GMR}$ to $M_{orb}$

In order to estimate the magnetic susceptibility of graphene, a geometrical model of the orbital current loop is needed. The easiest model one can imagine is the one of a current flowing along the edges of the gated region of graphene. This is equivalent to a thin rectangular loop carrying the orbital current. The edges parallel to the GMRs mostly contribute to the detected magnetic field. From Biot-Savart law we can determine the horizontal component of the magnetic field detected by the GMR. This value at the point C, the center of the GMR, is computed from the distances from C to the parallel edges of the gated region : $d_1 = 1.75$ $\mu$m and $d_2 = 5.57$ $\mu$m as well as the angles $\theta_{h1} = 30.96^o$, $\theta_{h2} = 9.3^o$ $\alpha_1 = 76^o$, $\alpha_1 = 51.5^o$ shown in Fig. S10.

$$B_{GMR} = \frac{\mu_0 I_{orb}}{2\pi} \left[ \frac{\sin(\theta_{h1})}{d_1} \sin(\alpha_1) - \frac{\sin(\theta_{h2})}{d_2} \sin(\alpha_2) \right] \tag{S.4}$$

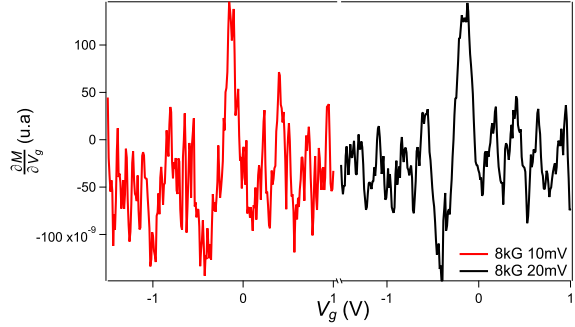

Figure S7: Comparison of the gate voltage derivative of the magnetisation measured with a 10 and 20mV modulation.

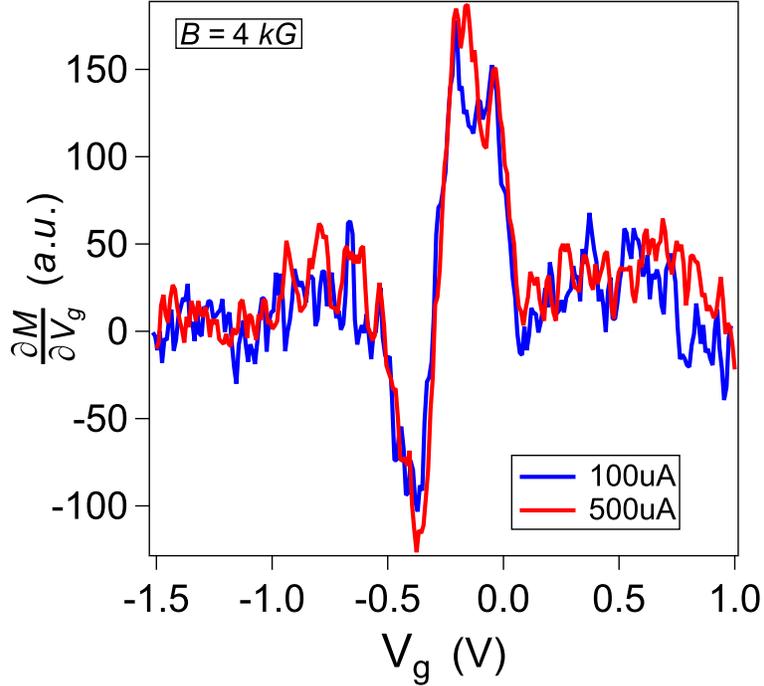

Figure S8: $\frac{\partial M}{\partial V_g}$ as a function of $V_g$ obtained for $B_{perp} = 4$ kG. The data measured with a current of $500\mu$ A and $100\mu$A through the GMR do not show any substantial difference.

from where we can find the coefficient relating the orbital current (or equivalently the magnetization per unit surface) to the field measured by the GMR sensor:

$$I_{orb}/B_{GMR} = M_{orb}/(SB_{GMR}) = 22.3(A/T) \qquad (S.5)$$

We also applied the same model to compare the experimental value of the magnetic field induced on the GMR by a rectangular gold loop (e-beam lithography and metal evaporation) which was deposited between the GMRs. The dimensions of the loop shown in Fig. S11were $4 \times 8\mu m$, corresponding to the right lithographic pattern. A ac modulation of te current through the loop was used to determined the response of the GMR to the field generated by the loop. A sensitivity of $4 \times 10^{-2}T/A$ was obtained.

In order to compare this result with our simple magnetostatic model, we approximate the loop by a infinitesimally thin rectangle . We consider that the GMR is only sensitive to the field generated by the parallel edges of the loop (the contribution of the perpendicular edges is neglected). Using again the Biot-Savart law it can be shown that the horizontal field at point P is given by equ. S.4. Taking into account the loop geometry we obtained the values of the angles $\alpha_1 = 65.7^o$ and $\alpha_1 = 35.53^o$, the other parameters being unchanged compared with graphene.

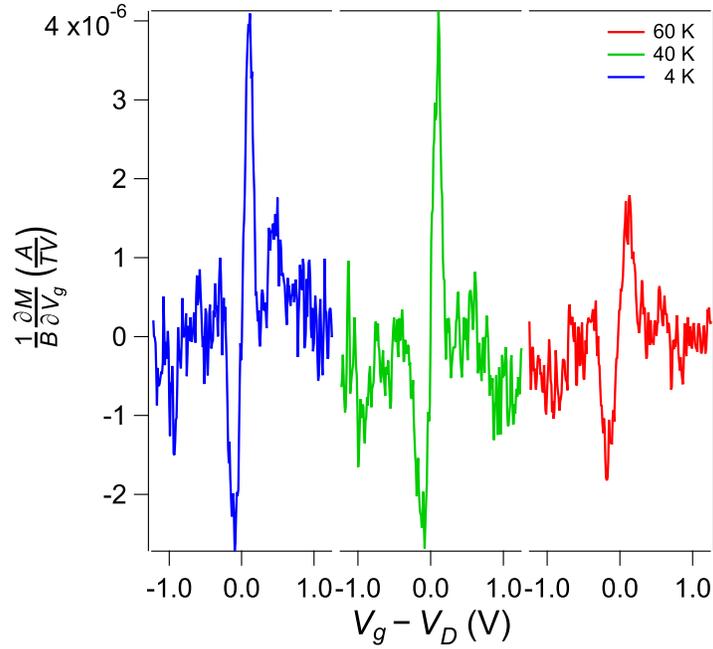

Figure S9: $\frac{1}{B}\frac{\partial M}{\partial V_g}$ as a function of $V_g$ for different values of temperature. One can notice that the peak amplitude does not vary between 4.2 and 40 K but is reduced at 60 K.

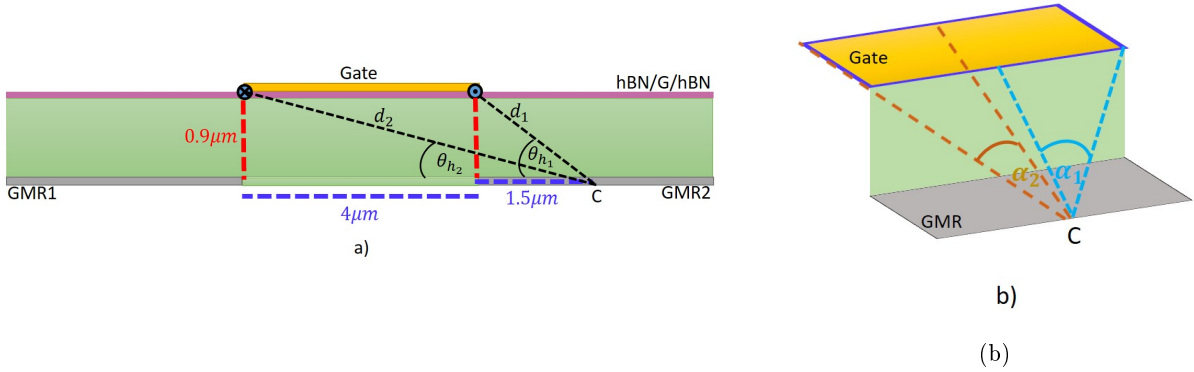

Figure S10: Front and lateral views of the sample with the definition of the angles $\theta_{h1}$ and $\theta_{h2}$ between the 2 edges of the gate electrode and the plane of the GMR. C is the center of symmetry of the GMR detector.

We obtain a value of $5.2 \times 10^{-2} T/A$, which agrees within 20% with the experimental value.

These numbers allow to quantitatively compare the experimental data measured on graphene with theoretical predictions. The analysis is made with the averaged data $(\partial M/\partial V_g)$ in order to decrease the errors introduced by integration. We therefore consider the amplitude $A_{d\chi_m}$ measured between the positive and negative peaks in $(1/B)(\partial M/\partial V_g)$, around $V_g = V_D$. We find $A_{d\chi_m} = 3.7 \pm 0.5 \mu A(VT)^{-1}$ for B=0.1T and $A_{d\chi_m} = 4.3 \pm 0.5 \mu A(VT)^{-1}$ for B=0.2T.

In the model with a gaussian distribution of $\mu$, we expect the following amplitude for the McClure susceptibility peak:

$$\chi_{Gauss}(0) = \frac{2e^2 v_F^2}{3\pi^{3/2}\sqrt{2}\sigma_0} = 0.95 \left[\frac{\mu A}{T}\right] \tag{S.6}$$

However, we cannot in principle directly compare our experimental data to the McClure peak, defined for vanishing external magnetic field. The correct procedure is to calculate the gate voltage dependent magnetization divided by the magnetic field within our theoretical model and compare its derivative with the gate voltage. We find for the theoretical equivalent of $A_{d\chi_m}$ defined above:

$$A_{d\chi_t} = 9.594 \times 10^{-6} \left[\frac{A}{VT}\right] \qquad (S.7)$$

which gives a ratio theory-experiment of 2.5 for the data taken at 0.1 T and 2.2 for the data taken at 0.2 T . We finally note that a better agreement between experiment and theory is obtained at larger field where the width of the McClure peak does not depend on $\sigma_0$ but only on $\epsilon_B$.

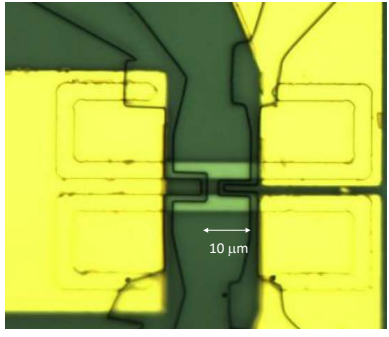

Figure S11: Lithography pattern of current loops fabricated to calibrate the GMRs sensors. Only the right one was operational.

### 7. Preliminary measurements on a strained sample

In the following we present measurements performed on a hBN/graphene/hBN stack deposited on a similar GMR detector with a 100 nm thick bottom gate below the 800 nm thick alumina layer. This stack compared to the stacks made using a top gate described in the main text is presumably highly strained. The signal detected on the GMR exhibits a peak which is much wider than in the experiment due to the much thicker dielectric substrate. One can see in Fig. S12 that the observed peak in $dR_{GMR}/dV_g$ is still present in zero magnetic field and does not change sign between 1000 and -1000 G. This intriguing result calls for additional experiments where strain is applied in a controlled way.

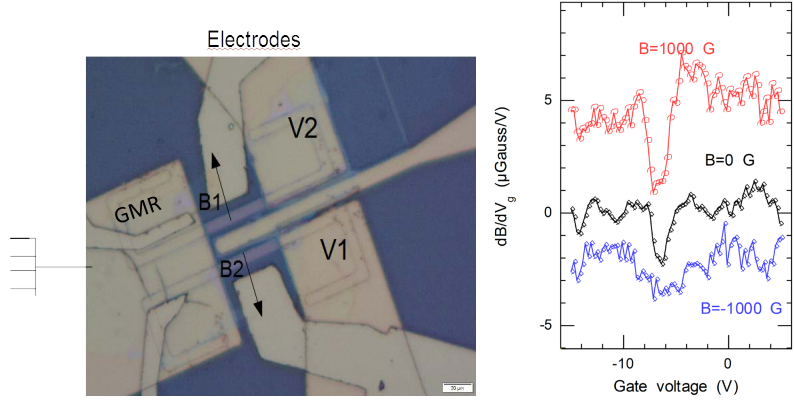

Figure S12: Investigation of a strained BN encapsulated graphene stack on a thick gate electrode (left panel). The signal obtained on the GMR sensor exhibits asymmetric peaks centered on the Dirac point with a sizable peak at zero field.

## II. SUPPLEMENTARY TEXT

### A. Theoretical calculations

The experiment described in the paper addresses the dependence of the magnetization $M$, more precisely the derivative $\partial M/\partial V_g$, as a function of the gate voltage $V_g$. In this supplemental material, we propose a theoretical derivation of this quantity. This is done in two steps, the dependence of the magnetization versus chemical potential $M(\mu)$ and the relation between $\mu$ and the gate voltage $V_g$. Special attention is given to the broadening due to a distribution of the electrochemical potential in the presence of disorder.

#### 1. Grand potential as a function of the chemical potential

Several expressions for the field dependent part of the grand potential in graphene are found in the literature, including the original paper by McClure[1, 12, 41, 42]. Here we propose the following derivation. The electronic spectrum in a field is written as

$$\epsilon_n = \pm v_F \sqrt{2|n|\hbar eB} \equiv \epsilon_B \sqrt{|n|} \tag{S.8}$$

with degeneracy $2eB/h$ per unit area, taking into account the spin degeneracy ($\epsilon_B^2 = 2\hbar v_F^2 eB$). The grand potential is a double integral of the density of states per unit of area $\nu(\epsilon, B)$ which is written:

$$\nu(\epsilon, B) = \frac{2eB}{h} \sum_{n,\pm} \delta \left( \epsilon \pm \epsilon_B \sqrt{|n|} \right) \ . \tag{S.9}$$

A Poisson transformation leads to the Fourier decomposition of the density of states :

$$\nu(\epsilon, B) = \frac{2|\epsilon|}{\pi \hbar^2 v_F^2} \left( 1 + 2 \sum_{p=1}^{\infty} \cos \frac{2\pi p \epsilon^2}{\epsilon_B^2} \right) \tag{S.10}$$

After a double integration, we obtain the oscillatory part of the grand potential for a clean sample and at zero temperature :

$$\Omega_0(\mu, B) = \frac{\epsilon_B^3}{4\pi^2 \hbar^2 v_F^2} \Delta_0 \left( \frac{\mu}{\epsilon_B} \right), \tag{S.11}$$

with

$$\Delta_0(x) = \sum_{p=1}^{\infty} \frac{1}{p^{3/2}} [1 - 2S(2\sqrt{p}|x|)] \tag{S.12}$$

where $S(x)$ is the Fresnel integral :

$$S(x) = \int_0^x \sin \frac{\pi t^2}{2} dt . \tag{S.13}$$

This variation, first obtained by McClure (although in a different form) is recalled in Fig. S14-a. On an energy scale larger than $\epsilon_B$, the function can be replaced by a $\delta$-function having the same total weight. The substitution $1 - 2S(|x|) \rightarrow \frac{4}{\pi}\delta(x)$ transforms eq. (S.11) into:

$$\Omega_0(\mu, B) = \frac{\epsilon_B^4}{12\pi\hbar^2 v_F^2}\delta(\mu) = \frac{e^2 v_F^2 B^2}{3\pi}\delta(\mu) \tag{S.14}$$

sometimes called the McClure peak.

At finite temperature, in the presence of elastic disorder, or with a distribution of the electrochemical potential around the average chemical potential (coinciding with Fermi energy at zero temperature), this expression has to be convoluted with one of the corresponding functions:

$$P_T(\epsilon) = \frac{\beta/4}{\cosh^2 \beta\epsilon/2} ,$$
$$P_D(\epsilon) = \frac{T_D}{\epsilon^2 + (\pi T_D)^2} ,$$
$$P_\sigma(\epsilon) = \frac{1}{\sqrt{2\pi}\sigma} e^{-\frac{\epsilon^2}{2\sigma^2}} . \tag{S.15}$$

Here we consider that the main source of broadening is due to the distribution $P_\sigma(\mu')$ for the electrochemical potential $\mu' = \mu_D - \langle\mu_D\rangle$ assumed to be Gaussian with a standard deviation $\sigma$. In graphene, the efficiency of the screening of charged impurities giving rise to the disorder potential increases indeed with doping, that is when moving away from the Dirac point. Therefore the distribution of $\mu'$ is expected to depend on $\mu$, so that the standard deviation $\sigma$ is a function $\sigma(\mu)$ which decreases with $|\mu|$, see Fig. S13.

Here, we present the calculation of the grand potential, with a fixed value of $\sigma$.

$$\Omega_\sigma(\mu, B) = \int_{-\infty}^{+\infty} P_\sigma(\mu')\Omega_0(\mu - \mu', B)d\mu'$$
$$= \frac{\epsilon_B^3}{4\pi^2 \hbar^2 v_F^2}\Delta_\sigma\left(\frac{\mu}{\epsilon_B}\right) \tag{S.16}$$

with

$$\Delta_\sigma(x) = \sum_{p=1}^{\infty} \frac{1}{p^{3/2}} \int_{-\infty}^{\infty} \frac{e^{-y^2}}{\sqrt{\pi}} [1 - 2S(2\sqrt{p}|x + \frac{\sqrt{2}\sigma}{\epsilon_B}y|)] dy \tag{S.17}$$

This function is plotted in Fig. S14 -b for $\sigma/\epsilon_B = 0.1$ which corresponds to $\sigma = 42$ K for $B = 1$ T.

In the limit $\sigma \gg \epsilon_B$, one recovers the Gaussian decay

$$\Delta_{\sigma \gg \epsilon_B}(\mu/\epsilon_B) = \frac{\pi\epsilon_B}{3}P_\sigma(\mu) \longrightarrow \Omega_{\sigma \gg \epsilon_B}(\mu, B) = \frac{e^2 v_F^2 B^2}{3\pi}P_\sigma(\mu) \tag{S.18}$$

or a decay as $P_T(\mu)$, $P_D(\mu)$ if temperature or elastic disorder are the main sources of broadening.

From the grand potential, we deduce the magnetization $M = -\partial\Omega/\partial B$ (here we compute $-\partial\Omega/\partial\epsilon_B$ noting that $\partial/\partial B = (\epsilon_B/2B)\partial/\partial\epsilon_B$). The dependence of this quantity versus chemical potential is displayed on Fig. S15. In principle all these calculations could also be done taking an explicit dependence of $\sigma(\mu)$.

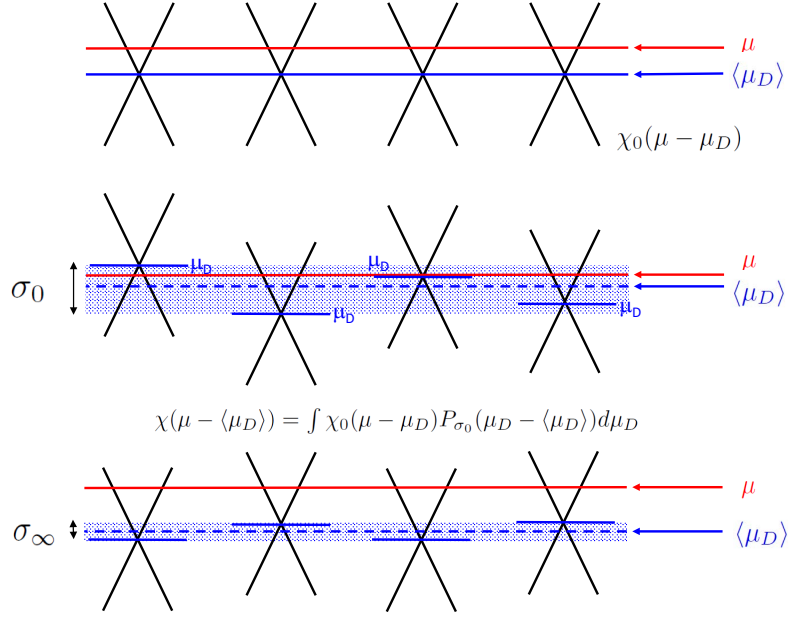

Figure S13: *Schematic representation of the fluctuation of electrochemical potential $\mu' = \mu_D - \langle\mu_D\rangle$, measured with respect to average Dirac point $\langle\mu_D\rangle$. They are induced by a screened disorder potential produced by charge impurities which amplitude decreases when $\mu - \langle\mu_D\rangle$ increases: The standard deviation $\sigma$ depends on $\mu$ and decreases when carrier density increases.*

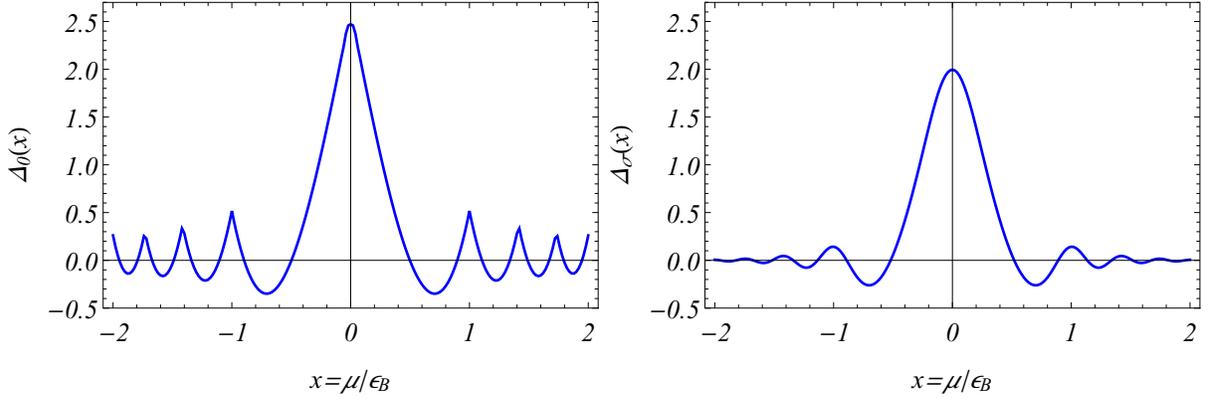

Figure S14: *Functions $\Delta_0(x)$, and $\Delta_\sigma(x)$ for $\sigma/\epsilon_B = 0.1$.*

### 2. Non-linear field dependence of the magnetization

We take the opportunity of the supplemental material to discuss the field dependence of the diamagnetic response, here at fixed chemical potential $\mu = 0$. In the presence of broadening, the field dependent part of the grand potential is given by :

$$\Omega(B) = \int P(\mu')\Omega_0(\mu', B)d\mu' \tag{S.19}$$

where $\Omega_0(\mu, B)$ and $P(\mu)$ are given by eqs. (S.11,S.12) and (S.15). Two limits are of special interest:

When the broadening is large, that is $T, T_D$ or $\sigma \gg \epsilon_B$,

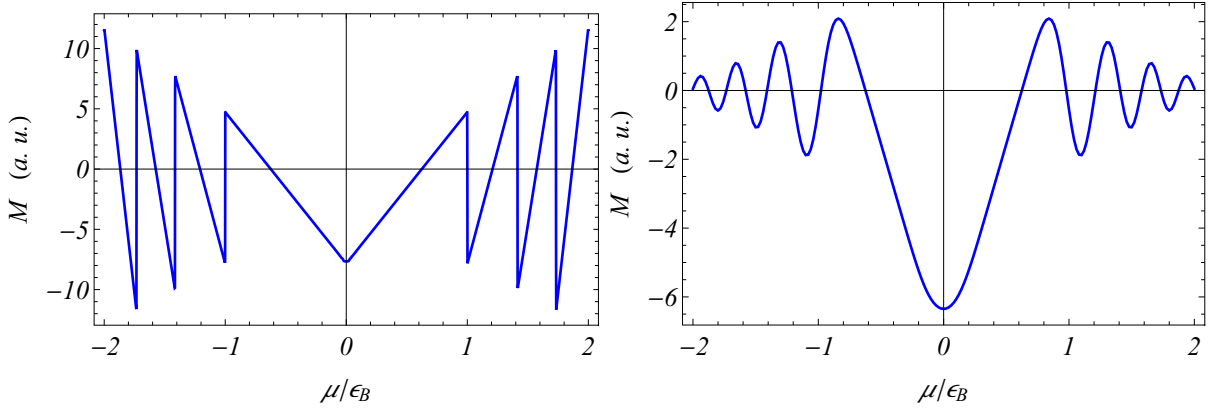

Figure S15 Magnetization as a function of $\mu/\epsilon_B$, for $\sigma = 0$ and $\sigma/\epsilon_B = 0.1$ independent of $\mu$.

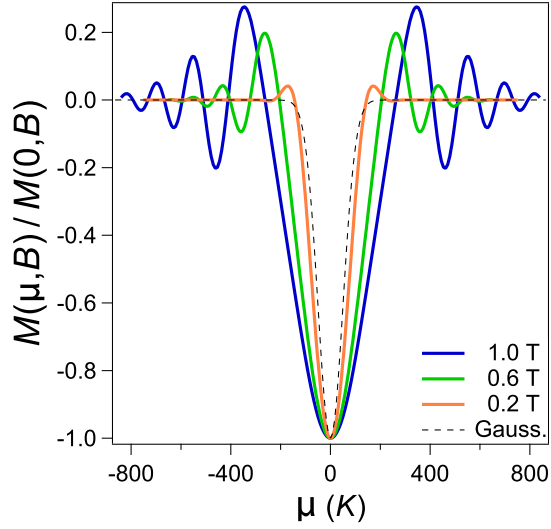

Figure S16: Theoretical magnetisation (normalized to its value at $\mu = 0$) as a function of $\mu$ expressed in Kelvin for $\sigma = 50$ K. The low field magnetisation is a gaussian of width sigma. The different plots show the broadening of the peak which increases as $\epsilon_B$, with increasing magnetic field and become independent on $\sigma$. On the other hand de Haas-von Alphen oscillations show up. Their damping is directly related to $\sigma$ and can be used to determine $\sigma$ at large doping.

$$\Omega(B) = P(\mu = 0) \int \Omega_0(\mu', B) d\mu' = \frac{e^2 v_F^2 B^2}{3\pi} P(\mu = 0) \tag{S.20}$$

leading to a quadratic field dependence of the grand potential as $B^2 \times \min(1/T, 1/T_D, 1/\sigma)$ and a magnetization linear in B. For the specific case of a Gaussian distribution of width $\sigma$, the grand potential reads in this limit:

$$\Omega_\sigma(B) = \frac{\sqrt{2}e^2 v_F^2 B^2}{6\pi^{3/2}\sigma} \tag{S.21}$$

In the opposite limit of a perfectly clean sample or very strong field, the field dependence becomes non-analytical [1] as :

$$\Omega(B) = \Omega_0(0, B) = \frac{\epsilon_B^3}{4\pi^2 \hbar^2 v_F^2} \zeta(3/2) = \frac{v_F e^{3/2} \zeta(3/2)}{\pi^2 \sqrt{2\hbar}} B^{3/2} \tag{S.22}$$

since $\Delta_0(0) = \zeta(3/2)$ and the magnetization is proportional to $\sqrt{(B)}$ . Note that all limits can be summarized as

$$\Omega(B) \propto B^2 \times \min(1/\epsilon_B, 1/T, 1/T_D, 1/\sigma) \tag{S.23}$$

The non-linear field dependence of the magnetization is difficult to observe [12]. Authors in [12] have investigated deviations from the linearity at moderate magnetic field. A description of the interpolating regime has been proposed by [12] using a Langevin function. We stress here that the correct behavior, see equation (S.19), deviates significantly form a Langevin function, in particular in small field.

### 3. Gate voltage $V_g(\mu)$

It is of fundamental importance to find the relation between $V_g$ and $\mu$ given that in our experiment the control variable is precisely the gate voltage. We start by modeling the action of $V_g$ as the one of a capacitance per unit surface relating $V_g$ to the charge density in graphene: $V_g \times C_g = en$:

$$V_g = \frac{en}{C_g} = \frac{e}{C_g \pi} k^2 = \alpha \operatorname{sign}(\mu - \mu_D)(\mu - \mu_D)^2, \tag{S.24}$$

with $\alpha = e/(C_g \pi \hbar^2 v_F^2)$. In the model of a Gaussian distribution of $\mu'$, this relation takes the following form, assuming that $C_g$ is the geometrical capacitance between graphene and the gate and therefore independent of $\mu$:

$$V_g(\mu) = \frac{\alpha}{\sqrt{2\pi}\sigma} \int_{-\infty}^{\infty} \operatorname{sign}(\mu - \mu')(\mu - \mu')^2 \exp\left(-\frac{\mu'^2}{2\sigma^2}\right) d\mu' \tag{S.25}$$

After integration, we get:

$$V_g(\mu) = \alpha \times \operatorname{erf}\left(\frac{\mu}{\sqrt{2}\sigma}\right)(\mu^2 + \sigma^2) + \frac{4\alpha}{\sqrt{2\pi}}\mu\sigma \times \exp\left(-\frac{\mu^2}{2\sigma^2}\right) \tag{S.26}$$

where erf is the error function: $\operatorname{erf}(x) = \frac{2}{\sqrt{\pi}} \int_0^x e^{-t^2} dt$.

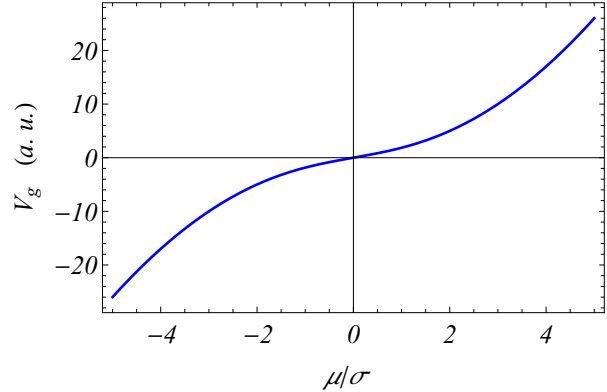

Figure S17: Relation between the gate voltage and the chemical potential, assuming that $\sigma$ is independent of $\mu$. Note the linear dependence of $V_g(\mu)$ at low $\mu$ compared to $\sigma$.

It is easy to generalize equation (S.26) to the case where $\sigma$ depends on $\mu$. It leads then to the two following expressions, respectively valid in the limits of low and large $\mu$ compared to $\sigma_0$:

$$V_g(\mu) = 4\sigma_0 \mu/\sqrt{2\pi} \text{ for } \mu \ll \sigma_0$$
$$V_g(\mu) = \alpha\mu^2 \operatorname{sign}(\mu) \text{ for } \mu \gg \sigma_0 \tag{S.27}$$

In graphene, the efficiency of the screening of charged impurities giving rise to the disorder potential increases with doping, that is when moving away from the Dirac point. Therefore the fluctuations of $\mu'$ are expected to depend on $\mu$: the standard deviation $\sigma$ should be a function $\sigma(\mu)$ which decreases with $|\mu|$. We denote $\sigma_0$ the value of $\sigma(\mu)$ close to the Dirac point and $\sigma_\infty$ its limiting value far from the Dirac point. Experimentally our data yields $\sigma_0 \simeq 165K$ which is much larger than $\sigma_\infty \simeq 50K$ determined from the damping of de Haas-van Alphen oscillations at large $\mu \gg \sigma_0$. This is due to the formation of electron-hole puddles in the vicinity of the Dirac point [43, 44].

In order to fit experimental data it was enough to consider only these two values. $\sigma_0$ determines the width of the low field McClure peak and the variation of $V_g(\mu)$ at low chemical potential. On the other hand $\sigma_\infty$ describes the large field damping of the de Haas-van alphen oscillations for $\mu \gg \sigma_0$. We note that in these large fields where de Haas-van alphen oscillations in $\mu$ are visible, the width of the magnetisation peak at $\mu = 0$ is determined by $\epsilon_B$ and does not depend on $\sigma$, which simplifies the fit of experimental data.